\documentclass[sigconf]{aamas}  
\pdfoutput=1
\usepackage{booktabs}
\usepackage{mathtools}
\usepackage{enumitem}

\setcopyright{ifaamas}  
\acmDOI{}  
\acmISBN{}  
\acmConference[AAMAS'18]{Proc.\@ of the 17th International Conference on Autonomous Agents and Multiagent Systems (AAMAS 2018)}{July 10--15, 2018}{Stockholm, Sweden}{M.~Dastani, G.~Sukthankar, E.~Andr\'{e}, S.~Koenig (eds.)}  
\acmYear{2018}  
\copyrightyear{2018}  
\acmPrice{}  

\newcommand{\ISN}{\mathsf{ISN}}
\newcommand{\CISN}{\mathcal{C-}\mathsf{ISN}}
\newcommand{\set}[1]{\{#1\}}	

\begin{document}

\title{Industrial Symbiotic Networks  as  Coordinated  Games}  
\subtitle{Extended Abstract}

\author{Vahid Yazdanpanah}
\affiliation{%
  \institution{University of Twente}
  \city{Enschede} 
  \country{The Netherlands} 
}
\email{V.Yazdanpanah@utwente.nl}

\author{Devrim Murat Yazan}
\affiliation{%
  \institution{University of Twente}
  \city{Enschede} 
  \country{The Netherlands} 
}
\email{D.M.Yazan@utwente.nl}

\author{Henk Zijm}
\affiliation{%
  \institution{University of Twente}
  \city{Enschede} 
  \country{The Netherlands} 
}
\email{W.H.M.Zijm@utwente.nl}

\begin{abstract}  
We present an approach for implementing a specific form of collaborative industrial practices---called Industrial Symbiotic Networks ($\ISN$s)---as MC-Net cooperative games and address the so called $\ISN$ implementation problem. This is, the characteristics of $\ISN$s may lead to inapplicability of \emph{fair} and \emph{stable} benefit allocation methods even if the collaboration is a collectively desired one. Inspired by realistic $\ISN$ scenarios and the literature on normative multi-agent systems, we consider \emph{regulations} and normative socioeconomic \emph{policies} as two elements that in combination with $\ISN$ games resolve the situation and result in the  concept of \emph{coordinated} $\ISN$s. 
\end{abstract}


\maketitle
 

\section{Introduction}

Industrial Symbiotic Networks ($\ISN$s) are  collaborative networks of industries with the aim to reduce their materials and energy footprint by circulating reusable resources (e.g, physical waste material) among the network members \cite{chertow2000industrial,lombardi2012redefining,yazan2016design}. Such a symbiosis leads to socioeconomic and  environmental benefits for involved firms and the society. One barrier against stable $\ISN$ implementations is the lack of  frameworks able to secure such networks against unfair and unstable allocation of obtainable benefits among the involved firms. In other words, even if economic benefits are foreseeable, lack of stability and/or fairness may lead to non-cooperative decisions and hence unimplementability of $\ISN$s ($\ISN$ \emph{implementation} problem). Reviewing recent contributions in the field of industrial symbiosis research, we encounter  studies focusing  on the interrelations between industrial enterprises \cite{yazan2016design} and the role of contracts in the process of $\ISN$ implementation \cite{albino2016exploring}. We believe a missed element for shifting from \emph{theoretical} $\ISN$ design to \emph{practical} $\ISN$ implementation  is to model, reason about, and support $\ISN$ decisions in a \emph{dynamic} way---and not by using snapshot-based modeling frameworks. 

This  abstract  reports  on  extending  the  game-theoretic approach  of \cite{iesm2017} with \emph{regulative} rules and normative socioeconomic \emph{policies}---following the successful line of  work on normative multi-agent systems \cite{shoham1995social,grossi2013norms,andrighetto2013normative}. The extension provides a scalable  solution to the $\ISN$ implementation problem and enables enforcing  desired industrial collaborations in a fair and stable manner.

\subsection{Research Questions}

The  following  questions  guide  the  design  of  a game-theoretic  framework and  its normative coordination mechanism that jointly facilitate the implementation of $\ISN$s:
\begin{enumerate}[leftmargin=\parindent]
\item $\ISN$ \emph{Games}: How to define a game-theoretic basis for $\ISN$s that both reflects their operational cost dynamics and allows the integration of normative rules?
\item $\ISN$ \emph{Coordination}: How to uniformly represent the regulatory dimension of $\ISN$s using incentive rules and normative policies?      
\item \emph{Coordinated} $\ISN$ \emph{Games}: How to develop a framework that integrates  normative coordination methods into $\ISN$ games to enable the fair and stable implementation of desirable $\ISN$s---with respect to an established policy?
\end{enumerate}
Dealing with $\ISN$s' complex industrial context \cite{DBLP:conf/eumas/YazdanpanahYZ16}, an ideal $\ISN$ implementation platform would be tunable to specific industrial settings, scalable for implementing various $\ISN$ topologies, and would not require industries to sacrifice financially nor  restrict their freedom in the market. Below, we present the overview of an approach for developing an $\ISN$ implementation framework with properties close to the ideal one.

\section{Overview of The Approach}

As discussed in \cite{albino2016exploring,iesm2017}, the total obtainable cost reduction (as an economic benefit) and its allocation among involved firms are key drivers behind the stability of $\ISN$s. For any set of agents involved in an $\ISN$, this value---i.e., the obtainable cost reduction---characterizes the value of the set and hence can be seen as a basis for formulating $\ISN$s as cooperative games.  On the other hand,  in realistic $\ISN$s, the symbiotic practice takes place in presence of economic, social, and environmental \emph{policies} and under \emph{regulations} that aim to enforce the policies by nudging  the behavior of agents towards desired ones. This is, while policies generally indicate whether an $\ISN$ is ``good (bad, or neutral)", the regulations are a set of norms that---in case of agents' compliance---result in an acceptable spectrum of collective behaviors. We follow this normative perspective and aim to use normative coordination to guarantee the implementability of desirable $\ISN$s---modeled as games---in a stable and fair manner. In the following subsections, we indicate how $\ISN$ games can be modeled and coordinated using regulatory incentive rules and normative socioeconomic policies.

\subsection{ISNs as Cooperative Games}

In the  game-theoretic representation of $\ISN$s, the value of any set of agents $S$ is defined \cite{iesm2017} using the difference between the total  cost that firms have to pay  in case the $\ISN$ does not occur, i.e. costs to discharge wastes and to purchase traditional primary inputs (denoted by $T(S)$), and the total cost that firms have to pay collectively in case the $\ISN$ is realized, i.e.  costs for recycling and  treatment, for transporting resources among firms, and  transaction costs (denoted by $O(S)$). Formally, the $\ISN$ among agents in a non-empty finite set of agents $N$ is a normalized superadditive cooperative game $(N,v)$ where for $S\subseteq N$,  $v(S)$ is equal to $T(S)- O(S)$ if $|S|>1$, and $0$ otherwise.  

Benefit sharing is crucial in the process of $\ISN$ implementation, mainly because of  stability and fairness concerns. Roughly speaking, firms are rational agents that defect unbeneficial collaborations  (instability) and mostly tend to reject relations in which benefits are not shared according  to contributions (unfairness). Focusing on the Core and Shapley allocations \cite{osborne1994course,mas1995microeconomic}---as standard methods that characterize stability and fairness---these solution concepts appear to be  applicable in a specific class of $\ISN$s but are not generally scalable for value allocation in the implementation phase of $\ISN$s. In particular, relying on the balancedness of two-person $\ISN$ games, denoted by $\ISN_{\Lambda}$, we can show  that any $\ISN_{\Lambda}$ is implementable in a fair and stable manner. However, in larger games---as balancedness does not hold necessarily---the core of the game may be empty which in turn avoids an $\ISN$ implementation that is reasonable for all the involved firms. So, even if a symbiosis could result in collective benefits, it may not last due to instable or unfair implementations. A natural response which is in-line with realistic practices is to employ monetary incentives as a means of normative coordination---to guarantee the implementability of ``desired'' $\ISN$s. To allow a smooth integration with normative rules, we transform $\ISN$ games into \emph{basic MC-Nets}\footnote{A basic MC-Net represents a  game in $N$ as a set of rules $\{\rho_i : (\mathcal{P}_i,\mathcal{N}_i) \mapsto v_i  \}_{i \in K}$, where $\mathcal{P}_i  \subseteq N$, $\mathcal{N}_i \subset N$, $\mathcal{P}_i \cap \mathcal{N}_i = \emptyset$, $v_i \in \mathbb{R} \setminus \{0\}$, and $K$ is the set of rule indices. For a group $S\subseteq N$, a rule $\rho_i$ is  applicable if $\mathcal{P}_i \subseteq S $ and $\mathcal{N}_i \cap S=\emptyset $. Then $v(S)$ will be equal to $\sum_{i \in \Pi(S)} v_i$ where $\Pi(S)$ denote the set of rule indices that are applicable to $S$. This rule-based representation allows natural integration with rule-based coordination methods and results in relatively low complexity for computing allocation methods such as the Shapley value \cite{lesca2017coalition,ieong2005marginal}.}  through the following steps: let $(N,v)$ be an arbitrary $\ISN$ game, $S_{\geq 2}= \{S \subseteq N : |S|\geq 2 \}$ be the set of all groups with two or more members where $K=|S_{\geq 2}|$ denotes its cardinality. We start with an empty set of MC-Net rules. Then for all  groups $S_{i} \in S_{\geq 2}$, for $i= 1$ to $K$,  we add a rule $\{\rho_i : (S_i,N \setminus S_i) \mapsto v_i=T(S_i)-O(S_i)  \}$ to the MC-Net. 

\subsection{Normative Coordination of ISNs}

Following \cite{shoham1995social,grossi2013norms}, we see that norms can be employed as  game transformations to bring about more desirable outcomes in $\ISN$ games. For this account,  given the economic, environmental, and social dimensions and with respect to potential socioeconomic consequences, $\ISN$s can be partitioned in three classes by a normative socioeconomic policy function $\wp : 2^N \mapsto \set{p^+,p^\circ , p^-}$, where $N$ is a finite set of firms. Moreover,  $p^+$, $p^\circ$, and  $p^-$ are labels---assigned by a third-party authority---indicating whether  an $\ISN$  is promoted, permitted, or prohibited, respectively. 

The rationale behind introducing policies  is mainly to make sure that the set of promoted $\ISN$s are implementable in a fair and stable manner while prohibited ones are instable. To ensure this, in real $\ISN$ practices, the regulatory agent  introduces monetary incentives, i.e., ascribes subsidies to promoted and taxes to prohibited collaborations. We follow this practice and employ a set of rules to ensure/avoid the implementability of desired/undesired $\ISN$s by allocating incentives\footnote{See \cite{DBLP:conf/ijcai/MeirRM11,DBLP:conf/atal/ZickPJ13} for similar approaches on incentivizing cooperative games.}. Such a set of incentive rules can be represented by an MC-Net $\Re = \{\rho_i : (\mathcal{P}_i,\mathcal{N}_i) \mapsto \iota_i  \}_{i \in K}$ in which $K$ is the set of rule  indices. Then, the incentive value for $S \subseteq N$,  is defined as $\iota(S)\coloneqq\sum_{i \in \Im(S)} \iota_i$ where $\Im(S)$ denotes the set of rule indices that are applicable to $S$.  It is provable that for any $\ISN$ game there exists a set of incentive rules to guarantee its  implementability.   

\subsection{Coordinated ISN Games}

Having policies and regulations, we integrate them into $\ISN$ games and  introduce the concept of \emph{Coordinated $\ISN$s} ($\CISN$s). Formally, let $G$ be an $\ISN$  and $\Re$ be a  set of regulatory incentive rules, both as MC-Nets among agents in $N$. Moreover, for each group $S\subseteq N$, let $v(S)$ and $\iota(S)$ denote the value of $S$ in $G$ and the incentive value of $S$ in $\Re$, respectively. We say the Coordinated $\ISN$ Game ($\CISN$) among agents in $N$ is a cooperative game $(N,c)$ where for each group $S$, we have that $c(S)=v(S) + \iota(S)$.

It can be observed  that employing such incentive rules is effective for  enforcing socioeconomic policies. In particular, we have that for any promoted $\ISN$ game, under a policy $\wp$, there exist an implementable  $\CISN$ game. Analogously, similar properties hold while \emph{avoiding} prohibited $\ISN$s  or \emph{allowing} permitted ones. The presented approach for incentivizing $\ISN$s is advisable when the policy-maker is aiming to ensure the implementability of a promoted $\ISN$ in an ad-hoc way. In other words, an $\Re$ that ensures the  implementability of a promoted $\ISN$ $G_1$ may ruin the implementability of another promoted $\ISN$ $G_2$. To avoid this,  the set of collaborations that a policy $\wp$ marks as promoted should be mutually exclusive. Accordingly, we have the desired result that the mutual exclusivity condition is sufficient for ensuring the implementability of \emph{all} the $\ISN$s among $\wp$-promoted groups  in a fair and stable manner.  

\section{Concluding Remarks}

The details of the components for developing the $\ISN$ implementation framework---rooted in cooperative games and coordinated with normative rules---consist of algorithms for generating incentive rules and policy properties to ensure the implementability of promoted $\ISN$s. We plan to explore the possibility of having multiple policies and tools for policy option analysis \cite{sara2017structured} in $\ISN$s. Then, possible regulation conflicts can be resolved using prioritized rule sets (inspired by formal argumentation theory \cite{modgil2013general,kaci2008preference}). We also aim to focus on administration of $\ISN$s by modeling them as normative multi-agent organizations \cite{boissier2013organisational,DBLP:conf/eumas/YazdanpanahYZ16} and relying on norm-aware frameworks \cite{dastani2016commitments,aldewereld2007operationalisation} that enable monitoring organizational behaviors. 

\begin{acks}
The project leading to this work has received funding from the European Union's Horizon 2020 research and innovation programme under grant agreement No. 680843. 
\end{acks}

\bibliographystyle{ACM-Reference-Format}  
\bibliography{References}

\end{document}